\documentclass[aps,prb,twocolumn,showpacs,superscriptaddress,citeautoscript]{revtex4}
\usepackage{amsmath}
\usepackage{amssymb}
\usepackage{graphicx}
\usepackage{dcolumn}
\usepackage{bm}
\renewcommand{\vec}[1]{\bm{#1}}
\usepackage{times}
\usepackage{array}
\begin{document}

\title{Stabilization of magnetic skyrmions by RKKY interactions}

\author {Alla V. Bezvershenko}
\affiliation{Physics Department,
Taras Shevchenko National University of Kyiv,  03022 Kyiv, Ukraine}

\author {Alexei K. Kolezhuk}
\affiliation{Institute of High Technologies,
Taras Shevchenko National University of Kyiv,  03022 Kyiv, Ukraine}
\affiliation{Institute of Magnetism, National Academy of Sciences
and Ministry of Education and Science,  03142 Kyiv, Ukraine}

\author {Boris A. Ivanov}
\affiliation{Institute of Magnetism, National Academy of Sciences
and Ministry of Education and Science,  03142 Kyiv, Ukraine}
\affiliation{National University of Science and Technology
``MISiS'', Moscow, 119049, Russian Federation}

\begin{abstract}
We study the stabilization of an isolated magnetic skyrmion in a magnetic
monolayer on non-magnetic conducting substrate via the
Ruderman-Kittel-Kasuya-Yosida (RKKY) exchange interaction. Two different types
of the substrate are considered, usual normal metal and single-layer graphene.
While the full stability analysis for skyrmions in the presence of the RKKY
coupling requires a separate effort that is outside the scope of this
work, we are able to study the radial stability (stability of a skyrmion against
collapse) using variational energy estimates obtained within the first-order
perturbation theory, with the unperturbed Hamiltonian describing the isotropic
Heisenberg magnet, and the two perturbations being the RKKY exchange and the
easy-axis anisotropy.

 We show that a proper treatment of the long-range nature of the RKKY
 interaction leads to a qualitatively different stabilization scenario compared
 to previous studies, where solitons were stabilized by the frustrated exchange
 coupling (leading to terms with the fourth power of the magnetization
 gradients) or by the Dzyaloshinskii-Moriya interaction (described by terms
 linear in the magnetization gradients).  In the case of a metallic substrate,
 the skyrmion stabilization is possible under restrictive conditions on the
 Fermi surface parameters, while in the case of a graphene substrate the
 stabilization is naturally achieved in several geometries with a
 lattice-matching of graphene and magnetic layer.

\end{abstract}

\maketitle

\section{Introduction and general remarks}
\label{sec:intro}

Topological defects in magnets, in particular a special type of
topologically nontrivial spin textures in quasi-two-dimensional magnets known as magnetic skyrmions,
have recently attracted a great deal of attention in the context of
developing new types of magnetic memory
\cite{FertReyrenCros17,FertCrosSampaio13,NagaosaTokura13,Romming+13,Koshibae+15}.
Magnetic skyrmions may have sizes down to a few nanometers,  are
easily moved by small electrical currents
\cite{Jonietz+10,Yu+12,Iwasaki+13,FertCrosSampaio13}, and thus are
considered as one of the promising routes to high-density spin-based
information storage and processing
\cite{Iwasaki+13,Kiselev+11,AnneBernand-Mantel17}.

The discussion of non-one-dimensional topological solitons centers around the problem of
stability for static solitons. There is a general no-go argument, known as the Hobart-Derrick
theorem \cite{Hobart63,Derrick64}, stating that in more than one
spatial dimension, stationary
localized soliton solutions
within a continuum model including only quadratic terms in gradients
of the order parameter are unstable.  (Note that this theorem does
not apply \cite{BPGalkina93} to topological solitons with infinite energy such as
hedgehogs, or Bloch points). In magnets with
a unit vector $\vec{n}$ being the order
parameter (normalized magnetization for
ferromagnets, or Neel vector for antiferromagnets), the typical energy
functional of a continuum model can be written as
\begin{equation}
\label{model1}
   W[\vec{n}]=\frac{S^2}{a^d}\int d^d(x)\Big\{\frac{1}{2}\mathcal{J}a^2
     (\nabla \vec{n})^2 +\mathcal{K}(1-n_{z}^{2})\Big\},
\end{equation}
 where $d$ is the spatial dimension, $a$ is the lattice constant, $S$ is the
atomic spin, the quantity $\mathcal{J}$ is of the order of the exchange
integral, and the last term describes the uniaxial magnetic
anisotropy, $\mathcal{K}$ being the anisotropy constant. It is
easy to see that for three-dimensional (3D) localized topological solitons
with a characteristic radial size  $R$, this energy takes the form
$W_{3\mathrm{D}}(R)= \zeta_1 \mathcal{J}S^{2}(R/a) + \zeta_2 \mathcal{K}S^{2}(R/a)^3$.  Here and
hereafter, $\zeta_{i}$ denote numerical factors of the order of
unity. The energy $W_{3\mathrm{D}}(R)$ has no minima at any finite
values of the soliton radius $R \neq 0$, and thus  no stable soliton
solutions with finite energy are present in the ``standard'' model
\eqref{model1}.

For Lorentz-invariant field theories, the only way to overcome this
problem is to add terms with higher powers of gradients to the energy
functional \eqref{model1}. This path has been taken in the famous
Skyrm model, where stable topological solitons of a bosonic field were
used to describe hadrons;\cite{Skyrm61,Skyrm62} this model gave the
name ``skyrmion'' used now for topological solitons in a wide class of non-linear
field theories.  In the physics of magnetism, terms of the form
$a^{4}\widetilde{\mathcal{J}} (\nabla^2 \vec{n})^2$ appear naturally
in the transition from the lattice spin model to continuum theory, as
a next term in the gradient expansion (see, e.g.,
Ref.\ \onlinecite{hubert-book}). In this case, the energy of a 3D
localized soliton can be estimated as
\[
 W_{3D}(R) = \zeta_1 \mathcal{J}S^{2}(R/a) + \zeta_2 \mathcal{K}S^{2}(R/a)^3 +  \zeta_3 \widetilde{\mathcal{J}}S^{2}(a/R),
\]
and this function can have a minimum at
finite $R_{3\mathrm{D}} \sim a \sqrt{\widetilde{\mathcal{J}}
  /\mathcal{J}}$,
provided that $\widetilde{\mathcal{J}}>0$. However,
the existence
of topological magnetic solitons stabilized via this scenario is
questionable. First, in a standard model with only nearest-neighbor spin
exchange interaction a \emph{negative} value of
$\widetilde{\mathcal{J}}$ is obtained, precluding
stabilization. Second, in extended
models with frustrated spin couplings (e.g., next-nearest-neighbor
exchange interactions of the opposite sign) it is indeed possible to find some region of parameters where the condition
$\widetilde{\mathcal{J}} >0$ is satisfied, but typically the two exchange constants $\widetilde{\mathcal{J}}$ and
$\mathcal{J}$ have the same order of magnitude. The latter fact means
that the soliton radius $R_{3\mathrm{D}}$ is generally of the
order of the lattice constant $a$, so the macroscopic
approximation, leading to the continuum model \eqref{model1} and
providing the grounds for the topological stability arguments, cannot
be trusted unless the couplings are fine-tuned 
to make $ \mathcal{J} \ll \widetilde{\mathcal{J}}$.

It is worth mentioning that for the 1D case the standard continuum
model \eqref{model1}
gives $W_{1D}(R)= \zeta_1 \mathcal{J}S^{2}(a/R) + \zeta_2 \mathcal{K}S^{2}(R/a)$, which  leads to stable 1D
solitons (kinks or domain walls) with the macroscopic thickness
$R_{1\mathrm{D}} \sim a \sqrt{J/K} \gg a$.
2D magnets, which are our primary focus, are in a sense an intermediate case with unique properties.
The same scale estimates, applied to
2D standard model, give $W_{2D}(R) = \zeta_1 \mathcal{J}S^{2} + \zeta_2 \mathcal{K}S^{2}(R/a)^2$. Thus,
the soliton energy depends on its radius $R$ only due to the magnetic
anisotropy, and in the purely isotropic case ($\mathcal{K}=0$) the soliton is
in a neutral equilibrium: its energy is independent on $R$ (in fact,
this reflects the more general property of scale invariance of
the model \eqref{model1} with $\mathcal{K}=0$ and $d=2$).

The exact skyrmion
solution of this kind was obtained by Belavin and Polyakov (BP)
\cite{BelavinPolyakov75} in their pioneering work for
two-dimensional isotropic Heisenberg ferromagnet (FM) within a
continuum model \eqref{model1} including only quadratic terms in
gradients. The BP solution for a skyrmion with the topological charge
$\nu$,  can be written in the form
\begin{eqnarray}
\label{BP-sol}
&& n_{x}+in_{y}=\sin\theta\,
e^{i\varphi},\quad n_{z}=\cos\theta,\\
&& \tan(\theta/2)=(R/r)^{|\nu|},\quad \varphi=\nu\arctan(y/x),\nonumber
 \end{eqnarray}
where $R$ plays the role of the skyrmion radius, and the skyrmion center is at the origin.
The BP skyrmion has a finite energy that does not depend on its size $R$:
\begin{equation}
\label{BP-energy}
   E_{\mathrm{BP}}=4\pi|\nu| \mathcal{J}S^2,
\end{equation}
 and the skyrmion stability is topologically
protected. However, as for any neutral equilibrium, the actual
stability conditions are extremely sensitive to weak perturbations of
the model.

The perfect scale invariance is immediately broken by magnetic
anisotropy: as a result, the presence of
the easy-axis anisotropy, however small it is, leads to a collapse of a skyrmion  after
its size diminishes to about a few lattice constants
\cite{Waldner86}. The magnetic anisotropy is an
unavoidable property of real magnets. Thus, again, skyrmion solutions
can only be stabilized by some additional
interactions, which would lead to the increase of their energy at $R\to 0$ and
protect them against collapse.

Several mechanisms of such a stabilization have been explored.
First, similar to the 3D case, adding the higher-order gradient term
$\widetilde{\mathcal{J}} (\nabla^2\vec{n})^2$
with $\widetilde{\mathcal{J}} > 0$ to the energy \eqref{model1}
leads to stabilization of the 2D
soliton.\cite{IvanovStephanovich86,IvStephZm90,AbanovPokrovsky98}
Frustrated exchange can provide
the proper sign of $\widetilde{\mathcal{J}}$, stabilizing skyrmion
solutions \cite{Waldner08,Okubo+12,LeonovMostovoy15,LinHayami16}.
Contrary to the 3D case, the soliton radius $R_{2\mathrm{D}}$
is \emph{macroscopically large} in 2D, even though
$\widetilde{\mathcal{J}} \sim \mathcal{J}$. Indeed, the
leading exchange term, quadratic in gradients, is scale-invariant and thus
its contribution to the soliton energy $E_{\mathrm{BP}}$ does not
depend on $R$, so the value of the radius is determined
by the competition of the magnetic anisotropy and the higher-gradient
terms.
Scaling arguments as
used above show that the soliton energy behaves as
\[
 W_{2D}(R)= E_{\mathrm{BP}}  + \zeta_2 \mathcal{K}(R/a)^2 +
 \zeta_{3} \widetilde{\mathcal{J}} (a/R)^{2},
\]
and under natural assumptions $\widetilde{\mathcal{J}} \sim
\mathcal{J} \gg K$  the soliton radius 
$R_{2\mathrm{D}} \sim a(\widetilde{\mathcal{J}}/\mathcal{K})^{1/4} \gg
a$ is much larger than the lattice constant. Moreover, the
``additional'' competing energies are small, $\widetilde{\mathcal{J}}
(a/R_{2\mathrm{D}})^{2} \sim \mathcal{K}R^2_{2\mathrm{D}} \sim (\mathcal{J}^3
\mathcal{K})^{1/4} \ll E_{\mathrm{BP}}$.

Condensed matter theories, in particular, magnetism, can provide
a wider class of interactions than  relativistic
field theories of high-energy physics, limited by fundamental symmetries
of the space-time. Dipolar
interaction is known to be a stabilizing factor for bubble-domains
in micron magnetic films. This interaction is important for
stabilization of so-called bubble skyrmions of nanometer size as
well (see Ref.\ \onlinecite{AnneBernand-Mantel17} and references therein).
Dzyaloshinskii-Moriya interaction (DMI), described in continuum theory
by the terms \emph{linear} in the gradients of the order parameter
(Lifshitz invariants), is naturally present in non-centrosymmetric
magnets and on interfaces because of lowering of the local
symmetry \cite{BogdanovRossler-surf}. It has been theoretically
predicted that the DMI can stabilize various skyrmion states, including ground-state
skyrmions \cite{BogdanovYabl89} with negative energy $E<0$, metastable
skyrmions \cite{IvStephZm90}
with the energy $E \sim E_{\mathrm{BP}}$, and
skyrmion lattices \cite{BogdanovHubert94}.  This stabilization mechanism is
most commonly used in current experiments concerning skyrmions.
 Though in most cases
stable skyrmion states are realized in the form of a
skyrmion lattice \cite{Muhlbauer+09,Yu+10,Heinze+11,Pfleiderer11},
a single-skyrmion ground state can be stabilized in spatially
restricted geometries such as nanodisks or
nanoribbons \cite{Sampaio+13,Buettner+15}. Isolated skyrmions and disordered
skyrmion arrays (skyrmion liquid) have been observed in recent
experiments \cite{Romming+13,Romming+15,Moreau+16,Woo+16,Tserkov+2016}.

An interesting possibility has been proposed long ago
by Abanov and Pokrovsky \cite{AbanovPokrovsky98}, who argued that
Ruderman-Kittel-Kasuya-Yosida (RKKY) exchange interaction (which can
be realized in  a magnetic film on top of a metallic substrate) is capable of stabilizing a
single skyrmion in the bulk (unrestricted) geometry. However, although
the RKKY interaction is long-range,
only its short-range part (up to the 2nd or 3rd
neighbors) is actually considered in
Ref.\ \onlinecite{AbanovPokrovsky98}. Thus, in fact, Abanov and
Pokrovsky  have studied not the effect of the RKKY
interaction, but rather the effect of frustrated exchange, which in
essence falls back to the model with fourth-order gradient term
described above.
At the same time, Kambersk{\'y} \textit{ et al.} \cite{Kambersky+99}
have shown that under certain
conditions on the Fermi wavevector the long-range part of the RKKY
exchange causes singular contributions to the spin stiffness of a ferromagnet,
effectively modifying the long-wavelength spin wave dispersion
law. One may expect that this effect should manifest itself  in the
skyrmion stability as well.

In the present work, we revisit the problem of skyrmions in presence of the
long-range RKKY interaction. We show that the RKKY interaction can
indeed stabilize an isolated skyrmion, but the actual physics of this
effect is drastically different from the simplified picture drawn from
the ``cut-off'' short-range version of RKKY exchange studied by Abanov and Pokrovsky \cite{AbanovPokrovsky98}.
Two scenarios are considered: a
magnetic monolayer on a metallic substrate, and a sandwich of
lattice-matched magnetic and graphene monolayers. We exploit
the aforementioned property of the energy of 2D solitons, namely,
that the main part of the energy is $R$-independent, and thus the
stabilization is driven by weak interactions which can be treated
perturbatively as small corrections. We show that a
proper treatment of the long-range nature of the ``true'' RKKY
exchange interaction results in a negative contribution to the skyrmion
energy of the type $-(R/a)^{\alpha}$ with $0< \alpha \leq 1$. Combined
with the contribution of the easy-axis anisotropy which is positive
and proportional to $(R/a)^{2}$, this provides a novel mechanism of
the skyrmion stabilization.
In the case of a metallic substrate,  we find that the skyrmion
stabilization is possible under restrictive requirements on the
Fermi surface parameters, and the RKKY contribution is
non-analytic ($\alpha=\frac{1}{2}$). In the case of a graphene
substrate $\alpha=1$,
the stabilization is naturally achieved in several geometries,
provided that one is able to overcome the experimental challenge of
preparing a lattice-matched ferromagnet-graphene interface.

The paper is organized as follows: in Sect.\ \ref{sec:model} we
outline the model and the approach to calculating the contribution of
the RKKY interaction to the skyrmion energy, in Sect.\ \ref{sec:metal}
we analyze the effect of RKKY interaction for skyrmions in ferro- and
antiferromagnets on a metallic substrate, Sect.\ \ref{sec:graphene} presents
the analysis for the case of ferromagnet on a single-layer graphene
substrate, and Sect.\ \ref{sec:discuss} contains a brief discussion
and summary.

\section{Model and method}
\label{sec:model}

Consider a 2D magnet described by the Hamiltonian
$\mathcal{H}=\mathcal{H}_{0}+\mathcal{H}_{RKKY} + \mathcal{H}_{a}$, where
\begin{eqnarray}
\label{ham}
&& \mathcal{H}_{0}=J\sum_{\langle\vec{r} \vec{r'} \rangle}
\vec{S}_{\vec{r}}\cdot \vec{S}_{\vec{r'}}, \quad
\mathcal{H}_{a}=-K\sum_{\vec{r}} (S_{\vec{r}}^{z})^{2}, \nonumber\\
&& \mathcal{H}_{\rm RKKY}=\frac{1}{2}J'\sum_{\vec{r},\vec{\Delta} } f(\vec{\Delta})
\vec{S}_{\vec{r}}\cdot \vec{S}_{\vec{r}+\vec{\Delta}}.
\end{eqnarray}
Here $\vec{S}_{\vec{r}}$ are spin-$S$ operators acting at sites $\vec{r}=(x,y)$ of a
2D bipartite lattice, the summation in $\mathcal{H}_{0}$ is over
nearest-neighbor pairs $\langle \vec{r} \vec{r'} \rangle$,
$J$ is the isotropic nearest-neighbor
Heisenberg exchange interaction, $K>0$ is the
easy-axis anisotropy constant, $J'$ sets the magnitude of the RKKY interaction,
and dimensionless function $f(\vec{\Delta})$ determines the dependence of this
interaction on the vector of distance $\vec{\Delta}$ between spins.  Throughout
the paper, the distance $a_{0}$ between nearest-neighbor lattice sites is set to unity.

 We treat spins classically, replacing  them by vectors of length
 $S$. In the absence of the anisotropy and of the RKKY interaction, the continuum
field solution for a skyrmion with the topological charge $\nu$ has
the form
\begin{equation}
\label{BP-sol1}
\vec{S}_{\vec{r}}=S \eta_{\vec{r}}\vec{n}(\vec{r}),
\end{equation}
where $\vec{n}(\vec{r})$ corresponds to the  BP solution \cite{BelavinPolyakov75} given by (\ref{BP-sol}).
In the ferromagnetic (FM) case ($J<0$), the factor $\eta_{\vec{r}}$ is trivial
($\eta_{\vec{r}}=1$), and in the antiferromagnetic (AFM) case ($J>0$),
it takes oscillating signs $\eta_{\vec{r}}=\pm1$ on two different sublattices.

In the absence of any additional interactions, the energy of the
skyrmion is given by (\ref{BP-energy}), where the parameter
$\mathcal{J}$ is proportional to the absolute value of the nearest-neighbor exchange
constant $J$ (in the case of a square lattice $\mathcal{J}=|J|$ , and
for a honeycomb lattice $\mathcal{J}=|J|/\sqrt{3}$; generally, for a
lattice with the coordination number $Z$ one obtains
$\mathcal{J}=Z|J|/(4\mathcal{A}_{0})$, where
$\mathcal{A}_{0}$ is the area per lattice site in units of $a_{0}^{2}$).  The expression
(\ref{BP-energy}) is valid in the continuum limit $R\gg1$, and is
independent of the skyrmion radius $R$ up to lattice corrections
coming from higher order gradient terms.

We assume that $K/|J|\ll 1$ and $|J'/J|\ll 1$. Then, to the first
order in those small parameters, the corresponding corrections to
the skyrmion energy can be obtained simply by calculating the value
of  weak perturbations  $\mathcal{H}_{a}$ and $\mathcal{H}_{\rm
RKKY}$ taken on the unperturbed skyrmion configuration
(\ref{BP-sol}):
\begin{eqnarray}
\label{corr-anis}
\Delta E_{a} &=& KS^{2}\sum_{\vec{r}} (1- n_{z}^{2}(\vec{r})) \\
\label{corr-rkky}
\Delta E_{\rm RKKY} &=& \frac{1}{2}J'S^{2}\sum_{\vec{r},\vec{\Delta} }
f(\vec{\Delta}) \eta_{\vec{\Delta}}  ( \vec{n}_{\vec{r}}\cdot \vec{n}_{\vec{r}+\vec{\Delta}} -1).
\end{eqnarray}

\subsection{Contribution of the magnetic anisotropy}

We will be interested in the
case of the lowest energy skyrmion solution with the unit topological
charge $|\nu|=1$.
There is a slight subtlety here as the sum (\ref{corr-anis}) resulting from the anisotropy correction is formally
divergent for $|\nu|=1$. However, the presence of a weak  easy-axis anisotropy
changes the power-law decay of the BP skyrmion solution to an exponential one  at
distances $r$ large compared to the characteristic  domain wall width
$\ell_{0}\sim(|J|/K)^{1/2}$, so the sum (\ref{corr-anis})  gets effectively cut off at
$r\sim\ell_{0}$ rendering  the resulting correction  logarithmically enhanced
but finite
\cite{VoronovIvanovKosevich}:
\begin{equation}
\label{dE-anis}
\Delta E_{a}\simeq 8\pi K S^{2} (R^{2}/\mathcal{A}_{0})\ln(\ell_{0}/R).
\end{equation}
The above expression is valid for small skyrmion radii $R\ll \ell_{0}$. For
skyrmions with $|\nu|\geq 2$ this convergence problem does not arise, and their
energy correction from the easy-axis anisotropy is simply proportional to
$R^{2}$.

In what follows, we do not write down the anisotropy
contribution to the skyrmion energy explicitly, but it is always
assumed that it is present and competes with the contribution from the
RKKY interaction, preventing the soliton from unlimited expansion.

\subsection{Contribution of the RKKY interaction}

The RKKY correction (\ref{corr-rkky}), after substituting the explicit form
(\ref{BP-sol}) of the $\nu=1$ BP solution, takes the following form:
\begin{equation}
\label{rkky-1}
\Delta E_{\rm RKKY} = -J'S^{2}R^{2}\sum_{\vec{\Delta}, \vec{r}}  \frac{\eta_{\vec{\Delta}}\Delta^{2}
f(\vec{\Delta})}{(r^{2}+R^{2}) [(\vec{r}+\vec{\Delta})^{2}+R^{2} ]}.
\end{equation}
Although the resulting lattice sum can be computed directly, it is instructive
to obtain some analytical estimates first.
While the summand in (\ref{rkky-1}) is generally an oscillating function of
$\vec{\Delta}$, its dependence on $\vec{r}$ is smooth; for that reason, it is a
good approximation to pass to the continuum and convert the sum over $\vec{r}$ into
an integral, keeping the sum over $\vec{\Delta}$ intact. Before doing so, it is
convenient to  exploit the central symmetry of the expression $\eta_{\vec{\Delta}}f(\vec{\Delta})$,
i.e., the fact that replacing vector $\vec{\Delta}$ by
$-\vec{\Delta}$ leaves it invariant,
and rewrite the RKKY contribution (\ref{rkky-1}) as
\begin{eqnarray}
\label{rkky-2}
&& \Delta E_{\rm RKKY} = -J'S^{2}R^{2}\sum_{\vec{\Delta}} \eta_{\vec{\Delta}}\Delta^{2}
f(\vec{\Delta}) \\
 && \qquad \times \sum_{\vec{r}} \frac{   (r^{2}+ R^{2}+
  \Delta^{2})}{(r^{2}+R^{2})
  [(r^{2}+R^{2}+\Delta^{2})^{2}-4(\vec{r}\cdot\vec{\Delta})^{2} ]}
.\nonumber
\end{eqnarray}
Passing to the continuum in the sum over $\vec{r}$ above, it is easy to obtain the
following  ``semi-continuum''  approximation for the RKKY correction:
\begin{eqnarray}
\label{rkky-semicont}
&& \Delta E_{\rm RKKY} = -J'S^{2}\frac{\pi R}{\mathcal{A}_{0}}\sum_{\vec{\Delta}} \eta_{\vec{\Delta}}\Delta
f(\vec{\Delta})  g(\Delta/R),\\
 && g(x)=\frac{1}{\sqrt{4+x^{2}}}\ln\left(
\frac{x(3+x^{2}) +(1+x^{2})\sqrt{4+x^{2}}}{\sqrt{4+x^{2}}-x}
\right).\nonumber
\end{eqnarray}
Expressions (\ref{rkky-2}) and (\ref{rkky-semicont})  will be the basis for our
further analysis.

\begin{figure}[tb]
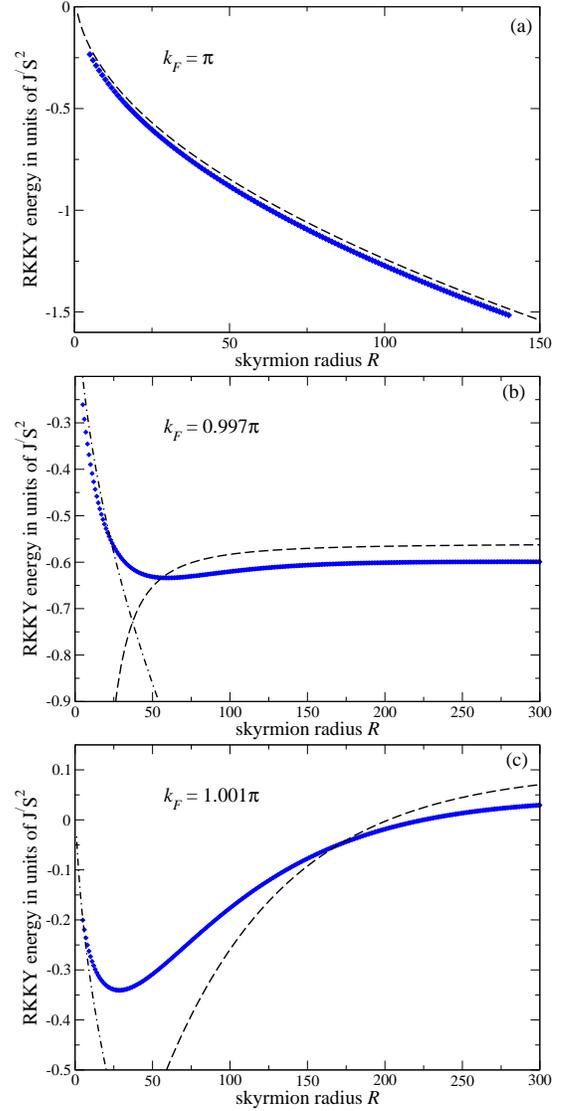

\includegraphics[width=0.4\textwidth]{compare_kF1}

\includegraphics[width=0.4\textwidth]{compare_kF0997}

\includegraphics[width=0.4\textwidth]{compare_kF1001}
\caption{\label{fig:compare}
 The RKKY contribution to the skyrmion energy as a function  of the skyrmion radius $R$ (the lattice
  constant is set to unity), for a square lattice, at different
  values of the Fermi wave vector $k_{F}$ around the first ``special
  point'': (a) $k_{F}=\pi$; (b) $k_{F}=0.997\pi$; (c) $k_{F}=1.001\pi$. Symbols denote the
  numerical results calculated via formula (\ref{rkky-semicont}), and lines correspond to asymptotic expressions
  (\ref{w-fm-asymp}). The cutoff in sum (\ref{rkky-semicont}) was
  set to $L=10^{4}$ for $k_{F}=\pi$, and to $L=2000$ for
  $k_{F}=0.997\pi$ and $k_{F}=1.001\pi$.
}
\end{figure}

\begin{figure}[tb]
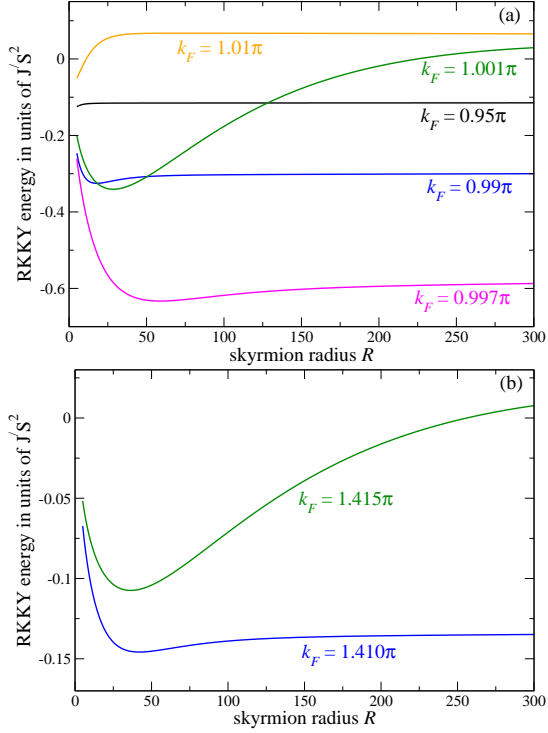

\includegraphics[width=0.4\textwidth]{deterio}

\includegraphics[width=0.4\textwidth]{deterio-kF141}

\caption{\label{fig:deterio}
 The RKKY contribution to the skyrmion energy as a function  of the skyrmion radius $R$ (the lattice
  constant is set to unity), for a square lattice, at different
  values of the Fermi wave vector $k_{F}$: (a) evolution of the $R$-dependence
  when moving away from the special value
  $k_{F}=\pi$; (b) behavior of the RKKY energy around the second
  special value $k_{F}=\pi\sqrt{2}$. The cutoff in the sum (\ref{rkky-semicont}) was
  set to $L=2000$.
 }
\end{figure}

\section{Metallic substrate}
\label{sec:metal}

For the case of a 2D magnet  on top of a metallic substrate,  we study the
effect of the RKKY interaction using the free
electron expression  \cite{RKKY}:
\begin{equation}
\label{metal-rkky}
f(\vec{\Delta})=\frac{\cos(2k_{F}\Delta)}{(2k_{F}\Delta)^{3}}-\frac{\sin(2k_{F}\Delta)}{(2k_{F}\Delta)^{4}},
\end{equation}
where $k_{F}$ is the Fermi wave vector (the radius of the Fermi sphere). For
real metals,  the situation is more complicated and will be discussed later.

It should be noted that Eq.\ (\ref{metal-rkky}) works fine even for
$\Delta \sim 1$ and thus can be taken as an adequate description of
both the  short- and long-range part of the RKKY exchange.

We will see that
the skyrmion stabilization takes place only in narrow regions around some
special values of $k_{F}$ which are different in the ferromagnetic and
antiferomagnetic cases. For the sake of simplicity, we consider the 2D
square lattice.

\begin{figure*}[tb]
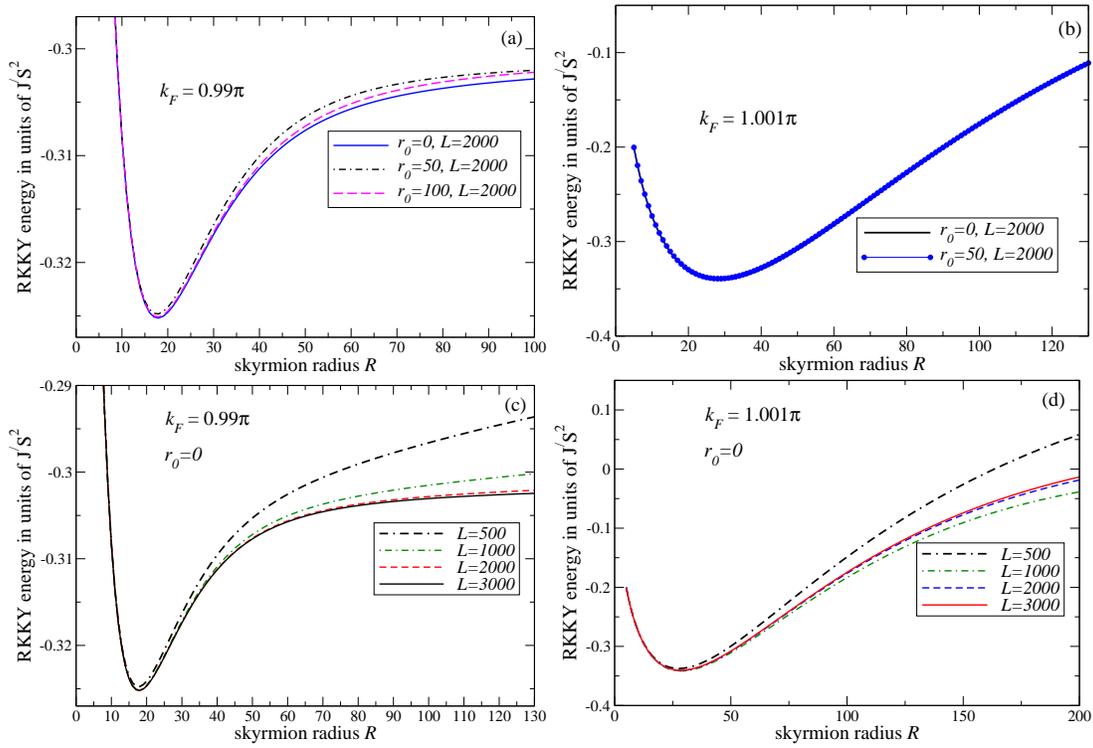

\includegraphics[width=0.4\textwidth]{kF099-pure-r0conv}
\includegraphics[width=0.4\textwidth]{kF1001-pure-r0conv}

\includegraphics[width=0.4\textwidth]{kF099-pure-Lconv}
\includegraphics[width=0.4\textwidth]{kF1001-pure-Lconv}

\caption{\label{fig:r0}
Analysis of errors in the RKKY contribution to the skyrmion energy, introduced by ``semi-continuum'' approximation and by
the finite cutoff $L$: (a,b)  results of the ``hybrid'' calculation using the
exact sum in (\ref{rkky-2}) over $|\vec{r}|$ smaller than a certain radius
$r_{0}$, and falling back to the continuum approximation in $\vec{r}$
for $|\vec{r}|>r_{0}$, for different values of $r_{0}$ at fixed
$L=2000$;  
(c,d) calculations in ``semi-continuum'' approximation ($r_{0}=0$) at
different values of the cutoff $L$.
 }
\end{figure*}

\subsection{Ferromagnet}
\label{subsec:fm-metal}

To obtain an analytical estimate for the dependence of the energy correction $\Delta E_{\rm RKKY}$ on the
skyrmion radius $R$, we will make a few simplifications.  First, we  drop
the second term in (\ref{metal-rkky}) as it decays much faster with distance
$\Delta$, and we will see that main contribution to the energy  comes from
$\Delta \sim R \gg 1$. Second, we observe  that  in
(\ref{rkky-semicont})  $\eta_{\vec{\Delta}}=1$ and the lattice sum over  $\Delta$
has the same structure as the sum computed by Kambersk{\'y} \textit{ et al.}
\cite{Kambersky+99} who studied the effect of the RKKY interaction on the spin
wave dispersion. Following Ref.\ \onlinecite{Kambersky+99}, we see that the
sum in (\ref{rkky-semicont}) has peculiarities if $2k_{F}$ is close to the
length of one of
the vectors $\vec{G}$ of the reciprocal lattice, i.e., if $k_{F}$ is close to one of the
``special points'' $\pi \sqrt{n^{2}+m^{2}}$,
where $n$ and $m$ are integers. Those are the same special points
around which a strong renormalization of spin wave stiffness occurs \cite{Kambersky+99}.

Let us consider the vicinity of  the first special point, setting
\begin{equation}
\label{fm-kf}
k_{F}=\pi(1-\delta)
\end{equation}
 with $|\delta| \ll 1$; then
the main
contribution to the lattice sum in (\ref{rkky-semicont}) comes from the four
``cones'' around the directions $\vec{\Delta}=(0,\pm 1)$
and $\vec{\Delta}=(\pm1,0)$.
After converting those sums to integrals,  one can easily analyze the
asymptotic behavior at small and large $R$ (see Appendix
\ref{app:integrals} for details). It is convenient to express the results in
terms of the quantity
\begin{equation}
\label{w-fm}
w_{\rm FM}=\frac{\pi^{2}(1-\delta)^{3}}{J'S^{2}}\Delta E_{\rm
  RKKY}\simeq -\sqrt{\frac{R}{8}} A(2\pi\delta R) +1,
\end{equation}
where the function $A(Q)$ is analyzed in Appendix
\ref{app:integrals} and has the following asymptotics:
\begin{eqnarray}
\label{A-asymp}
&& A(Q)\simeq 3.76 +16.47\, Q, \quad |Q| \ll 1, \nonumber\\
&& A(Q)\simeq
\begin{cases}
\sqrt{2\pi/Q} \Big(1+\frac{1}{8Q^{2}}\Big), & Q\gg 1\\
\frac{\pi}{2}\sqrt{\pi/|Q|} e^{-2|Q|}, & Q\ll -1
\end{cases}.
\end{eqnarray}

Then for small $R$ one obtains
\begin{subequations}
\label{w-fm-asymp}
\begin{equation}
\label{w-fm-1}
w_{\rm FM}\simeq
1-\sqrt{R}\left(1.33 + 5.82\cdot(2\pi\delta R) \right),\quad  2\pi\delta R\ll 1,
\end{equation}
while for large radii, $2\pi\delta R\gg 1$, the result is
\begin{equation}
\label{w-fm-2}
w_{\rm FM}\simeq
\begin{cases}
1-\frac{1}{\sqrt{8\delta}}\left(1 + \frac{1}{8(2\pi\delta R)^{2}}  \right),
 & \delta >0,\\
1-\frac{\pi}{8\sqrt{|\delta|}}e^{-4\pi|\delta| R}, & \delta<0
\end{cases} .
\end{equation}
\end{subequations}

One can see that the RKKY correction depends on the skyrmion radius $R$ in a
non-monotonic way, with a minimum. At intermediate $1\ll R\ll
(2\pi\delta)^{-1}$, the RKKY interaction to the
skyrmion is proportional to $-J'\sqrt{R}$, which, together with the
easy-axis anisotropy contribution (\ref{dE-anis}) that goes roughly as
$K R^{2}$, leads to  stabilization of the skyrmion radius at some value
$R_{0}\propto (J'/K)^{2/3}$. A similar picture can be obtained in the vicinity
of  $k_{F}=\pi\sqrt{2}$ and other ``special points''.

Fig.\ \ref{fig:compare} shows the comparison of numerical
results obtained by direct calculation of the ``semi-continuum'' sum
(\ref{rkky-semicont}) on a square lattice  with the asymptotic expressions
(\ref{w-fm-asymp}). One can see that indeed around the special value
  of  the Fermi wave vector $k_{F}=\pi$
the RKKY contribution tends to stabilize the skyrmion, and behaves
non-monotonically as a function of the skyrmion radius, in a good agreement
with the found asymptotics.
However, this stabilization effect deteriorates rapidly
away from the special value of $k_{F}$, as one can see in
Fig.\ \ref{fig:deterio}(a): when $k_{F}$ moves away from $\pi$, the minimum  moves
toward $R=0$ and disappears, and  $R$-dependence
becomes ``flat''.  Behavior of the RKKY energy around the second
special value $k_{F}=\pi\sqrt{2}$ is qualitatively the same as at
the first special point $k_{F}=\pi$, as shown in
Fig.\ \ref{fig:deterio}(b).

Approximating the discrete sum over $\vec{r}$ in Eq.\ (\ref{rkky-2}) by an integral
in the ``semi-continuum'' approximation of Eq.\ (\ref{rkky-semicont}) introduces an error which should be at greatest near the
skyrmion center, where one has the largest gradients. In order to
estimate this error, we have done ``hybrid'' calculations using the
exact sum (\ref{rkky-2}) for $|\vec{r}|$ smaller than some radius
$r_{0}$, and falling back to the continuum approximation in $\vec{r}$
for $|\vec{r}|>r_{0}$ (see Eq.\ (\ref{hybrid}) in Appendix
\ref{app:hybrid}); thus, $r_{0}=0$ is equivalent to
``semi-continuum'' approximation of Eq.\ (\ref{rkky-semicont}), and
$r_{0}\to\infty$ amounts to using Eq.\  (\ref{rkky-2}). 
Calculations for several values of $r_{0}$ are shown in
Fig.\ \ref{fig:r0}(a,b); one can see that the
semi-continuum approximation  works very well.

Another source of error is our use of a finite cutoff
$|\vec{\Delta}|<L$ in the sum (\ref{rkky-semicont}).
Fig.\ \ref{fig:r0}(c,d) presents  results of calculations for several
values of $L$ which show that for skyrmion radii $R<300$ the
convergence is already reached at $L=2000$.

\begin{figure}[tb]
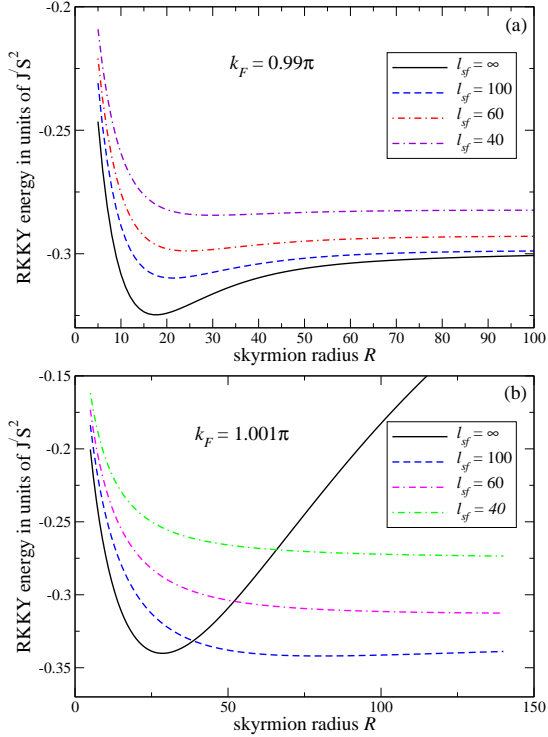


\includegraphics[width=0.4\textwidth]{kF099-sf}

\includegraphics[width=0.4\textwidth]{kF1001-sf}

\caption{\label{fig:sf}
  Influence of the finite spin-flip length
  $\ell_{sf}$ on the RKKY
  contribution to the skyrmion energy: (a) $k_{F}=0.99\pi$; (b)
  $k_{F}=1.001\pi$.
The cutoff in the sum over $\vec{\Delta}$ in (\ref{rkky-semicont}) was
  set to $L=2000$.
 }
\end{figure}

Further, we have studied how the skyrmion stabilization might be
affected by a finite spin-flip length $\ell_{sf}$ (the average
length an electron travels in metal before its spin gets flipped).  At
room temperature, in pure metals $\ell_{sf}$ ranges from tens to
several hundreds of nanometers, but it is strongly dependent on
temperature and is affected by the presence of interfaces
\cite{BassPratt07}.  One can expect that the RKKY interaction is
exponentially suppressed at distances larger than $\ell_{sf}$. We have
modeled this effect by simply introducing the factor
$\exp(-\Delta/\ell_{sf})$ into the RKKY power-law
(\ref{metal-rkky}). The results of such modification are shown in
Fig.\ \ref{fig:sf}. One can see that decreasing $\ell_{sf}$ leads to a
rapid ``flattening'' of the RKKY energy in the region of large
$R\gtrsim \ell_{sf} $, but leaves the initial square-root behavior
intact. Thus, even for relatively short $\ell_{sf}$ the
sum of the RKKY energy and the anisotropy contribution (\ref{dE-anis})
still has a minimum at some value of $R$ proportional to
$(J'/K)^{2/3}$, so the stabilization effect is preserved.

\subsection{Antiferromagnet}
\label{subsec:afm-metal}
\begin{figure}[tb]

\includegraphics[width=0.4\textwidth]{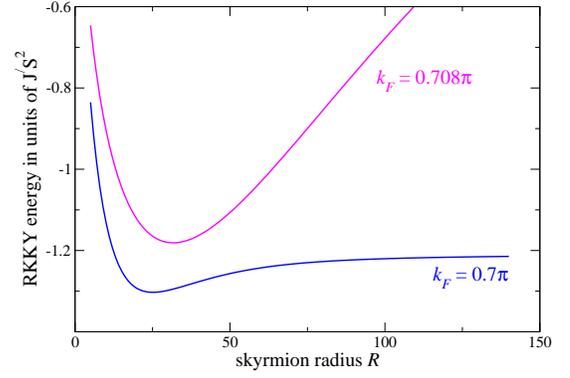}

\caption{\label{fig:afm}
  The RKKY
  contribution to the skyrmion energy for an antiferromagnet on a
  square lattice, around the first special point $k_{F}=\pi/\sqrt{2}$.
The cutoff in the sum over $\vec{\Delta}$ is $L=2000$.
 }
\end{figure}

In the antiferromagnetic case, the procedure is very similar to that
considered above. We again start from Eq.\  (\ref{rkky-semicont})
with $f(\vec{\Delta})$ given by the first term of Eq.\
(\ref{metal-rkky}), but  now we have the oscillating factor
$\eta_{\vec{\Delta}}=(-1)^{\Delta_{x}+\Delta_{y}}$, which changes
the ``special points'' in $k_{F}$ around which  the skyrmion
stabilization is possible: now $2k_{F}$ should be close to the
length of the vector $\vec{G}+(\pi,\pi)$, where
$\vec{G}=2\pi(n_{1},n_{2})$ is an arbitrary reciprocal lattice
vector. Thus the first ``special point'' is $k_{F}=\pi/\sqrt{2}$. As
shown in Appendix \ref{app:integrals}, one can obtain the analytical
estimate expressing the RKKY contribution to skyrmion energy via the
same function $A(Q)$:
\begin{equation}
\label{w-afm}
w_{\rm AFM}\simeq -\frac{\sqrt{R}}{2^{7/4}}A(Q=\pi\sqrt{2}\delta R) +1,
\end{equation}
where $w_{\rm AFM}$ is defined in full analogy to
Eq.\ (\ref{w-fm}). Thus, for the AFM case one can expect qualitatively
the same behavior of the RKKY energy as disussed above for the
ferromagnet. Numerical results confirm this prediction, as shown in
Fig.\ \ref{fig:afm}.

\subsection{Self-consistency conditions}
\label{subsec:selfcon}

As mentioned before, our calculations of the skyrmion energy can be viewed as a variational energy estimate obtained by
the first-order perturbation theory, with the unperturbed Hamiltonian
describing the isotropic Heisenberg magnet, and the two perturbations
being the RKKY exchange (proportional to $J'$) and the easy-axis
anisotropy (proportional to $K$). We look at the sector with the
topological charge $\nu=1$, where the unperturbed solution
is given by the BP soliton, and this solution is used 
as a basis to calculate contributions to the skyrmion energy from the two perturbations. 

Here, however, there is a subtle point: although, as we have shown, the RKKY
contribution is not singular in $J'$, it is well known that the contribution from the
easy-axis anisotropy is non-analytical in $K$ (see
Eq.\ (\ref{dE-anis})). This is due
to the fact that the anisotropy changes the power-law decay of the BP
solution to the exponential one at distances $r> \ell_0 \sim
(J/K)^{1/2}$, and the  BP
solution modified by the anisotropy is well described, e.g., by the
following ansatz proposed
in Ref.\ \onlinecite{VoronovIvanovKosevich})
\[
 \tan(\theta/2)=\frac{2}{(\nu-1)!} \left(\frac{R}{2\ell_{0}}\right)^{|\nu|}K_{\nu}\left(\frac{r}{\ell_{0}}\right),
\]
where $K_{\nu}(z)$ is the modified Bessel function of the second kind
(the Macdonald function). The above ansatz in principle could be used
for a direct computation of the double sum (\ref{corr-rkky}), 
but the integration over $\vec{r}$ could not be performed analytically. 
However, one can use the following  argument: 
the effect of the finite anisotropy on the RKKY contribution is
roughly equivalent to introducing cutoffs of about $\ell_0$  both for $r$
and $\Delta$ in the sum defined by Eq.\ (\ref{corr-rkky}). One can reasonably assume
that those cutoffs will not qualitatively affect our results for
skyrmion radii much smaller than the cutoff, $R \ll \ell_0$.
As we have shown above, for $k_F$ 
sufficiently close to one of the "special points"  the RKKY energy
gain is proportional to $R^{1/2}$ for $R\ll 1/(2\pi\delta)$, and then the
skyrmion energy minimum is reached at $R=R_0\sim (J'/K)^{2/3}$. 
Thus, our results should remain valid as long as $R_0$ is
much smaller than both the cutoff $\ell_0$ and  $1/(2\pi\delta)$,
which leads to the following
self-consistency conditions:
\[
J' \ll (J^{3}K)^{1/4},\quad J' \ll K (2\pi\delta)^{-3/2}.
\]

\section{Graphene substrate}
\label{sec:graphene}

The RKKY interaction in graphene is in many respects considerably
different from such in a normal two-dimensional metal (see, e.g.,
Ref.\ \onlinecite{PowerFerreira13} for a review). RKKY
interaction between magnetic adatoms at a graphene monolayer strongly
depends on how are the adatoms placed with respect to the carbons of
the graphene.  The geometry of the lowest energy configuration depends
on the adatoms concentration: for instance, while single Co atoms are
adsorbed on graphene ``hollow sites'' (centers of hexagons)
\cite{Donati+13}, graphene grown on Co(0001) substrate exhibits 3-fold
symmetry consistent with top-fcc or top-hcp configurations
\cite{Kaul+12,Usachov+15} where Co atoms of the proximate layer sit on
top of the carbons belonging to only one sublattice. This might be
caused by the van der Waals interaction contributing significantly to
the interaction between the substrate (or adatoms) and graphene
\cite{Stradi+11,Kaul+12}.

We focus on a simplified model of two lattice-matched
monolayers (graphene and magnetic). We will restrict ourselves to the case of ferromagnetic Heisenberg
exchange between magnetic atoms, and we will consider two possible
geometries: (a) magnetic atoms sitting on top of each carbon of the
graphene layer, forming a hexagonal lattice, and (b) magnetic atoms
sitting at the centers of hexagons (``hollow sites''), forming a
triangular lattice.

\subsection{``On-top'' configuration}
\label{subsec:top}

\begin{figure}[tb]

\includegraphics[width=0.38\textwidth]{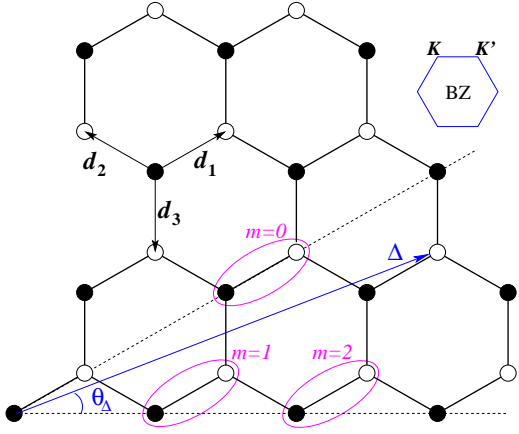}

\caption{\label{fig:graphene1}
  Schematic view of the graphene layer showing the
  notation used in Eqs.\ (\ref{ontop-aa}), (\ref{ontop-ab}). Two different
  sublattices are shown with full and open circles. The orientation of
  the Brillouin zone and the chosen adjacent pair of Dirac points
  $\vec{K}$, $\vec{K'}$ are indicated, as well as the three
  nearest-neighbor vectors $\vec{d}_{1,2,3}$. Three ellipses show the
  cluster of sites used when calculating the average (\ref{fav}).
 }
\end{figure}

In the case of magnetic atoms sitting on top of the carbon
atoms of undoped (half-filled) graphene, particle-hole symmetry leads
to ferromagnetic interaction between spins on the same graphene sublattice and
antiferromagnetic interaction of spins on different sublattices
\cite{Saremi07,Brey+07,Black-Schaffer10}.
In a simple tight-binding model of graphene with two  ``impurity''
spins sitting on top of carbons and coupled to the itinerant
electrons of graphene via the Heisenberg exchange $\tilde{J}$,  the
characteristic magnitude of the RKKY exchange is  \cite{Kogan11}
 $J'=a_{0}\tilde{J}^{2}/(256\hbar v_{F})$, where $v_{F}$ is the Fermi velocity and  $a_{0}$ is the carbon-carbon distance which
is hereafter set to unity.
The RKKY exchange $J'f_{AA,AB}(\vec{\Delta})$ between two impurity spins sitting on the
same sublattice (AA) or different sublattices (AB), where $\vec{\Delta}$ is the vector
connecting the two magnetic atoms, is determined
by  the following expresions
\cite{SherafatiSatpathy11,Kogan11}:
\begin{eqnarray}
\label{ontop-aa}
f_{AA}(\vec{\Delta}) &=& - \frac{1 + \cos[(\vec{K} - \vec{K'}) \cdot
    \vec{\Delta}]}{\Delta^3},\\
\label{ontop-ab}
f_{AB}(\vec{\Delta}) &=& 3 \frac{1 + \cos[(\vec{K} - \vec{K'}) \cdot \vec{\Delta} +\pi -2\theta_{\Delta} ]}{\Delta^3},
\end{eqnarray}
where
$\vec{K}$ and $\vec{K'}$ are a pair of adjacent Dirac points,
and $\theta_{\Delta}$
is the angle between vectors $\vec{\Delta}$ and $\vec{K'}-\vec{K}$ (see
Fig.\ \ref{fig:graphene1}). We choose
$\vec{K},\vec{K'}=\frac{2\pi}{3}(\mp \frac{1}{\sqrt{3}},1)$, then
Eqs.\ (\ref{ontop-aa}), (\ref{ontop-ab}) are valid  in the $60^{\circ}$-sector  $\theta_{\Delta}\in
[-\frac{\pi}{6},\frac{\pi}{6}]$; in the rest of the plane the pattern of
$f_{AA,AB}(\vec{\Delta})$ repeats according to the symmetry.
Two important points are worth noting: (i) although the RKKY interaction
(\ref{ontop-aa}), (\ref{ontop-ab}) is oscillating, both $f_{AA}$ and
$f_{AB}$ are not sign-changing; (ii) the magnitudes of
intra-sublattice and inter-sublattice interactions are different.

To analyze the RKKY contribution to skyrmion energy, we again use the
``semi-continuum'' approximation (\ref{rkky-semicont}),  where now the
sum over $\vec{\Delta}$ has to be broken into AA and AB parts, and the
area per lattice site is $\mathcal{A}_{0}=3\sqrt{3}/4$ (in units of $a_{0}^{2}$).
To deal
with oscillating contribution from the cosines in (\ref{ontop-aa}),
(\ref{ontop-ab}), it is
convenient to parametrize the distance vector as
$\vec{\Delta}=m(\vec{d}_{1}-\vec{d}_{2})+l\vec{d}_{1}$, where $m$ and
$l$ are non-negative integers ($l=1+3s$ for AB interaction and
$l=3s$ for the AA part, where $s=0,1,2,\ldots$), see
Fig.\ \ref{fig:graphene1}.  It is easy
to see that the pattern of $f_{AA}$ and $f_{AB}$ exhibits threefold periodicity
in $m$ as shown in Table \ref{tab:fun}.  Further,
$g(\Delta/R)$ in (\ref{rkky-semicont}) is a smooth function, and for $R\gg 1$
it changes only slightly when $\Delta$ changes by something of the
order of unity. Then, when computing the sum over $\vec{\Delta}$, we
can separate it into clusters of six sites with approximately the same $\Delta$,
as shown by three ellipses in Fig.\  \ref{fig:graphene1}. The value of $f(\vec{\Delta})$,  averaged
over such a six-site cluster, is
\begin{equation}
\label{fav}
\bar{f}(\vec{\Delta})=\frac{1}{6}
\sum_{j=0,1,2}\Big(f_{AA}^{m=j+3k}+f_{AB}^{m=j+3k} \Big) \simeq \frac{1}{\Delta^{3}},
\end{equation}
where we have just neglected differences in $m$ and $l$ between the
sites of the cluster. As one can see, this procedure is roughly equivalent
to neglecting the oscillating contribution of the cosines in
(\ref{ontop-aa}), (\ref{ontop-ab}). Because of unequal magnitudes of intra-sublattice and inter-sublattice
interactions, the averaged RKKY interaction (\ref{fav}) has positive
(AFM) sign, which is crucial for the skyrmion stabilization effect.

\begin{table}[tb]
\begin{tabular}{c|c|c|c}
\hline
 & $m=3k$ & $m=1+3k$ & $m=2+3k$ \\ \hline
$f_{AA}$ & $-\frac{2}{\Delta^{3}}$ & $-\frac{1}{2\Delta^{3}}$ &
$-\frac{1}{2\Delta^{3}}$ \\ \hline
$f_{AB}$ & $ \frac{6}{\Delta^{3}}-\frac{9m^{2}}{2\Delta^{5}}$  &
 $\frac{3l^{2}}{2\Delta^{5}}$ &
$\frac{3}{\Delta^{3}}+\frac{9m^{2}-3l^{2}}{2\Delta^{5}}$ \\ \hline
\end{tabular}
\caption{\label{tab:fun} Pattern of the RKKY interactions
  (\ref{ontop-aa}), (\ref{ontop-ab}); $m$, $l$, and $k$ are integer numbers,
  $\vec{\Delta}=(m+l)\vec{d}_{1}-m\vec{d}_{2}$, $\Delta^{2}=l^{2}+3m(m+l)$.}
\end{table}

After substituting the averaged RKKY amplitude (\ref{fav}) into
(\ref{rkky-semicont}), we can pass to continuum, converting the sum
over $\vec{\Delta}$ into an integral; this yields the following
asymptotic behavior of the RKKY contribution $\Delta E_{\text{RKKY}}$
to skyrmion energy:
\begin{eqnarray}
\label{ontop-asymp}
\frac{\Delta E_{\text{RKKY}}}{J'S^{2}} &\simeq&
-\frac{2\pi^{2}R}{\mathcal{A}_{0}^{2}}\int_{1/R}^{\infty}
\frac{dx}{x}g(x) \nonumber\\
&=& -\frac{16\pi^{4}}{27} R+\frac{32\pi^{2}}{27} +O(R^{-2}).
\end{eqnarray}
Thus, RKKY interaction via single-layer graphene leads to the
contribution into the skyrmion energy proportional to $-CJ'R$, with a
strong enhancement factor $C\sim 60$. Together with the
easy-axis anisotropy contribution (\ref{dE-anis}) that is roughly
proportional to $K R^{2}$, this leads to skyrmion stabilization  at
radius value $R_{0}\propto CJ'/K$.

\begin{figure}[tb]
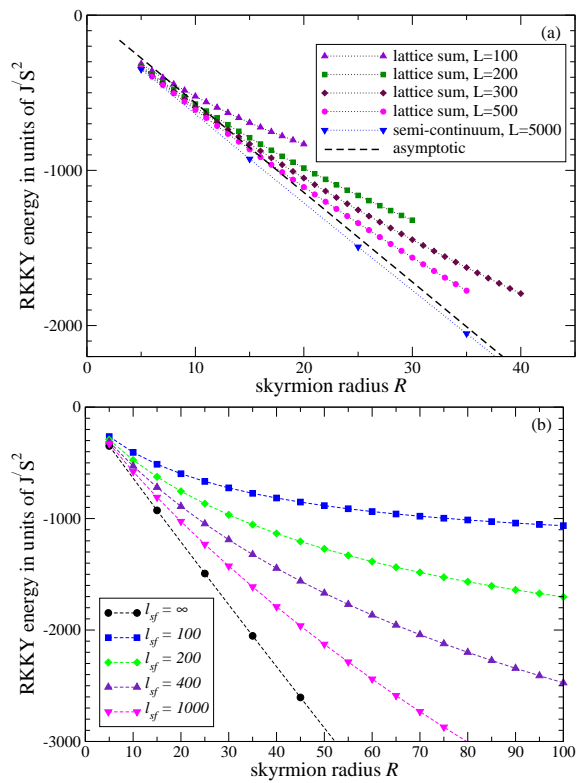


\includegraphics[width=0.42\textwidth]{graphene-cf}

\includegraphics[width=0.42\textwidth]{graphene-lsf}

\caption{\label{fig:graphene-cf}
  Numerical
results for the RKKY contribution to skyrmion energy, in a 2D
ferromagnet at single layer graphene in ``on-top'' geometry: (a) results obtained by direct calculation of the lattice sum
(\ref{rkky-2}) with different cutoffs (the same cutoff $L$ is used for
$|\vec{\Delta}|$ and $|\vec{r}|$) are shown in comparison to those obtained
 in semi-continuum approximation
(\ref{rkky-semicont}), the dashed line corresponds to the asymptotic formula
(\ref{ontop-asymp}); (b) the effect of a finite spin-flip length
 $\ell_{sf}$ (calculated in the semi-continuum approximation).
 }
\end{figure}

Fig.\ \ref{fig:graphene-cf}(a) shows the comparison of numerical
results obtained by direct calculation of the lattice sum
(\ref{rkky-2}) with several different cutoffs, and by using the
 ``semi-continuum'' sum
(\ref{rkky-semicont}), with the asymptotic expression
(\ref{ontop-asymp}).

Within the same approximation, it is easy to estimate the effect of a finite spin-flip
length as well. Introducing the exponentially decaying factor
$\exp(-\Delta/\ell_{sf})$ into (\ref{ontop-aa}), (\ref{ontop-ab}) as
we have done for the metallic substrate, we obtain
\begin{equation}
\label{ontop-asymp-lsf}
\frac{\Delta E_{\text{RKKY}}}{J'S^{2}} \simeq
-\frac{2\pi^{2}R}{\mathcal{A}_{0}^{2}}\int_{1/R}^{\infty}
\frac{dx}{x}g(x) \exp(-xR/\ell_{sf}),
\end{equation}
which falls back to (\ref{ontop-asymp}) for $R\ll \ell_{sf}$, and
goes to a constant, ${\Delta E_{\text{RKKY}}}/{J'S^{2}}\simeq
-\frac{32\pi^{2}R}{27}\ell_{sf}$ for $R\gg \ell_{sf}$.
The  influence of a finite spin-flip length is illustrated in Fig.\ \ref{fig:graphene-cf}(b).
The typical value of spin-flip length in undoped graphene \cite{Dlubak+12,Yan+16} is rather
large, $\ell_{sf}\sim 100$~$\mu$m which corresponds to about $10^{5}
a$, so this effect is likely to be negligible in practice.

\subsection{``Hollow sites'' configuration}
\label{subsec:hollow}

For magnetic atoms placed  at the ``hollow sites'' of graphene lattice
(i.e., at centers of hexagons), the RKKY interaction takes a very simple
form which is radially symmetric, non-oscillating, and
antiferromagnetic \cite{Saremi07,Black-Schaffer10,SherafatiSatpathy11}:
\begin{equation}
\label{hollow}
f(\vec{\Delta})=36 J'/\Delta^{3}.
\end{equation}
Calculation of the RKKY contribution to skyrmion energy in the
``semi-continuum'' approximation (\ref{rkky-semicont}) is thus
essentially reproducing the steps done above in the derivation of
Eq.\ (\ref{ontop-asymp}); there is just
an extra overall factor of $36$, and the area per magnetic atom is
doubled, $\mathcal{A}_{0}=3\sqrt{3}/2$. As a result, one obtains the
following asymptotic expression:
\begin{equation}
\label{hollow-asymp}
\frac{\Delta E_{\text{RKKY}}}{J'S^{2}} \simeq
 -\frac{16\pi^{4}}{3} R+\frac{32\pi^{2}}{3} +O(R^{-2}).
\end{equation}
Note that the enhancement factor (the number in front of $R$ in the
above formula) is in this case quite large, $\sim 500$. In other
respect, the effect of the RKKY interaction is in this case the same as for the
``on-top'' configuration.

Comparing the above results with those of Sect.\ \ref{sec:metal}, one
can notice that  the energy gain
due to the RKKY interaction is considerably larger for a
graphene substrate than for a normal metal. 
Indeed, according to the asymptotic formula (\ref{w-fm-2}), for a metallic substrate the maximal
energy gain in units of $J'S^2$ is limited by some value of the order
of unity, as illustrated by Fig.\ \ref{fig:compare}. At the same time, for a graphene substrate the RKKY energy gain is not capped and
continues to increase linearly with the skyrmion size $R$, as  follows from
the asymptotic formula (\ref{ontop-asymp}) and seen in
Fig.\ \ref{fig:graphene-cf}. 
The reason is, for a normal metal the
RKKY interaction oscillates in such a way that its net effect
integrates almost to zero, and the total energy gain is 
modest. For graphene, the situation is different (see
Eqs. (\ref{ontop-aa}), (\ref{ontop-ab})): (i)
either within the sublattice or between sublattices, RKKY oscillations
occur on top of a finite  value, in such a way that the sign of
the RKKY exchange stays constant; (ii) the magnitude of the
intersublattice RKKY exchange is three times larger than that of the
intra-sublattice one.  As a result, the net effect of the RKKY exchange in
graphene on the
skyrmion energy is much larger than in the case of a normal metal.

\section{Discussion and Summary}
\label{sec:discuss}

We have demonstrated the stabilizing effect of the RKKY interaction on
an isolated magnetic skyrmion in a 2D magnet, in two scenarios
involving the RKKY interaction via a normal metallic substrate or via
single-layer graphene. For both scenarios, we have used simple expressions
for the RKKY exchange obtained in models of non-interacting itinerant
electrons (free electrons with a spherical Fermi surface in the case
of a normal metal, and the tight-binding model for graphene). In both
cases, we have found that the RKKY interaction yields a negative contribution to the skyrmion
energy of the type $-(R/a)^{\alpha}$, $\alpha\leq1$, where $R$ is the
skyrmion radius. This contribution counteracts the skyrmion tendency
to collapse (caused by the contribution from the easy-axis anisotropy
proportional to $(R/a)^{2}$) and provides a novel mechanism of
the skyrmion stabilization.

For the RKKY coupling via a metallic substrate,
our main conclusion is
that the skyrmion stabilization is possible
under certain conditions on the Fermi wave vector $k_{F}$
(namely, in the case of a ferromagnet $2k_{F}$ should be close to the
length of one of the vectors $\vec{G}$ of the reciprocal lattice, and
for an antiferromagnet $2k_{F}$ should be close to the length of the
vector $\vec{G}+(\pi,\pi)$). The special values of $k_{F}$ are
exactly those where pecularities of the spin wave stiffness occur
\cite{Kambersky+99}, leading to modifications of the long-wavelength
spin wave dispersion that is non-analytic in the wave vector.
We have shown that if the above conditions on $k_{F}$ are satisfied, the
contribution of the RKKY interaction to the skyrmion
energy is proportional
to $-(R/a)^{1/2}$, non-analytic in $R$. Such non-analytic contributions, either to the
magnon spectrum or to the skyrmion energy, cannot be described by a
finite number of exchange interactions beyond the nearest neighbors,
so the resulting physics is very different from that studied by Abanov
and Pokrovsky \cite{AbanovPokrovsky98}.

One may
wonder whether the above-mentioned conditions on $k_{F}$ can  be realized in practice.
Extending the simple RKKY model to take into account real shapes of
Fermi surfaces\cite{Roth+66} leads to the RKKY oscillations containing,
instead of one wave vector $2k_{F}$,
several wave vectors $2k_{v}$ corresponding to the diameters
connecting so-called caliper points of the
Fermi surface.
As shown
in Ref.\ \onlinecite{Kambersky+99},
for the ideal epitaxial monolayer of magnetic atoms on a (001) plane of a
fcc metal with fcc lattice (Ag, Cu, Au), one has pairs of such caliper
points satisfying $2k_{v}=(1\pm\delta)|\vec{G}|$ with $\delta\approx 0.04\div0.06$.
Though being close to the ``resonance'', this might be still too far
from it for the stabilization, as seen from Fig.\ \ref{fig:deterio}.
Further fine-tuning of the shape of the Fermi surface might be
achieved by
diluting the metallic layer
\cite{SatoToth62,TempletonColeridge75}, or by
applying external
pressure \cite{Lazarev+65,Egorov+84,Gaidukov+84} (possibly also via the
interface-induced strain); the external
pressure can alter not only the shape, but even the topology of the Fermi surface \cite{Lazarev+65}.

For the RKKY coupling via a single-layer graphene, we have shown that
at least in two considered geometries (magnetic atoms on top of the
carbons on both sublattices, and magnetic atoms at hollow sites) the
stabilization is naturally achieved without any fine-tuning (only the
ferromagnetic case has been considered).  The crucial point, leading
to that effect, is the unequal magnitude of intra-sublattice
(ferromagnetic) and inter-sublattice (antiferromagnetic) RKKY
interaction in graphene: the inter-sublattice coupling is three times
stronger, which, on the net, leads to a negative (stabilizing) contribution to the
skyrmion energy proportional to the skyrmion radius $R$.  One may
speculate, that in the case of a lattice mismatch between the magnetic
layer and graphene, the average effect will be the same,
leading to skyrmion stabilization.  Actually, the only geometry
unfavorable for the skyrmion stabilization is that with magnetic atoms
on top of the only one carbon sublattice: then the RKKY contribution
to the  energy  is positive
and does not prevent the skyrmion collapse.

Experimentally, one of the problems of dealing with graphene/ferromagnetic interfaces
is the transfer of outer electrons of magnetic adatoms to the
graphene layer, which might lead to doping or to the presence of
weakly localized states in graphene, modifying the RKKY interaction \cite{DedkovFonin10,DedkovVoloshina15}.
Doping the graphene layer (or applying a gate voltage)
introduces additional oscillations with the finite Fermi wave vector $2k_{F}$ in
the RKKY interaction \cite{SherafatiSatpathy11a}, which  opens an
interesting possibility of tuning the $k_{F}$ to one of the singular points.

It should be emphasized that the approximate analysis used in the
present work can be viewed as a variational energy estimate obtained
within the first-order perturbation theory, with the unperturbed
Hamiltonian describing the isotropic Heisenberg magnet, and the two
perturbations being the RKKY exchange and the easy-axis
anisotropy. Although one can assess the stability of a skyrmion
against collapse in this way, it does not constitute a full proof of
stability. 
To analyze such a stability, one would need to find the full set of
magnon eigenmodes on top of the soliton background. While this problem can
be solved for the ``standard'' model with the local exchange
interaction \cite{Sheka+01}, the
eigenvalue problem in presence of long-range exchange
interactions is much more involved and remains, to the best of our
knowledge, unexplored. 
Nevertheless, one can make the following argument: the eigenmode
frequencies $\omega_{n,m}$ are
characterized by the azimuthal number $m$ and the principal number
$n$. An instability is reflected in the appearance of negative
$\omega_{n,m}^2$ for some values of $m$. The radial stability is
connected to $m=0$, and is fully determined by the energy $W(R)$
studied in the present work, because $\omega_{n,m=0}^2 \propto
d^2W(R)/dR^2$. 
Thus, a minimum in $W(R)$, as found in this work,
ensures that the soliton is stable against a radial perturbation,
i.e., against the collapse. Note that the only terms contributing to
the $R$-dependence of the energy are those breaking the
scale invariance, i.e., the RKKY coupling and anisotropy (but not the
nearest-neighbor exchange).
The $m=1$ mode describes a shift of the skyrmion as a whole and thus
does not correspond to any instability. 
A negative $\omega_{n,m=2}^{2}$ would correspond to the elliptical instability, which is, e.g., known
for magnetic bubbles in presence of the dipole-dipole
interaction. 
However, in the
absence of other interactions the stiffness coefficients
$\omega_{n,m}^{2}$ for $m\geq2$ are all positive\cite{Sheka+01} and their scale is
set by the nearest-neighbor
exchange $J$. Thus, one can expect that no instabilities arise if the
RKKY exchange and anisotropy are small compared to the
nearest-neighbor exchange, $J',K \ll J$ (as assumed throughout the
present paper). 
We hope that our analysis will stimulate the corresponding
numerical work which will deliver the true skyrmion solutions in
presence of the RKKY exchange and the easy-axis
anisotropy.

Finally, we would like to mention that we have considered only the effect of 2nd-order (RKKY)
exchange between spins of a magnetic insulator,
interacting  via the metallic substrate. In the case of itinerant
magnets, higher-order (multispin) interactions might
 become important \cite{Ozawa+17prl,*Hayami+17prb}, but this is
beyond the scope of the present work.

\begin{acknowledgments}

This work is supported by the National Academy of Sciences of Ukraine
via project No.~1/17-N.  One of us (B.A.I.) gratefully acknowledges
financial support of the Ministry of Education and Science of the
Russian Federation in the framework of Increase Competitiveness
Program of NUST ``MISiS'' (2-2017-005), implemented by the governmental
decree dated 16th of March 2013, No.~211.
\end{acknowledgments}

\appendix

\section{Asymptotic behavior of the RKKY energy correction}
\label{app:integrals}

In this Appendix we provide details of the analysis leading to the asymptotic
expressions (\ref{A-asymp}).  In all calculations, we assume that the skyrmion radius
is always much larger than the lattice constant, $R\gg 1$.

\paragraph{Ferromagnetic case.--}

We start from Eq.\  (\ref{rkky-semicont}) with $\eta_{\vec{\Delta}}=1$,
$f(\vec{\Delta})$ given by the first term of Eq.\ (\ref{metal-rkky}), and
$k_{F}=\pi(1-\delta)$, $|\delta|\ll 1$
and notice that the main
contribution to the lattice sum in (\ref{rkky-semicont}) comes from the four equivalent
``cones'' around the directions $\vec{\Delta}=(0,\pm 1)$,
$\vec{\Delta}=(\pm1,0)$. Following the procedure of
Ref.\ \onlinecite{Kambersky+99},  a contribution of one
``cone'' can be  approximately calculated as follows:  put $\vec{\Delta}=(n,m)$,
where $n$ and $m$ are integers,
with $|n| \ll m $, then inside the cosine $\cos(2k_{F}\Delta)$ one can expand $\Delta$ in $n$,
$\Delta\simeq m+n^{2}/(2m)$,
and in the rest of the expression one can set $\Delta\simeq m$, so  we have
\begin{eqnarray}
\label{ex1}
&& w_{\rm FM} \simeq -\frac{ R}{2} \sum_{m=1}^{\infty}  \frac{g(\frac{m}{R})}{m^{2}}
 \sum_{n=-m+1}^{m}
\Big\{
\cos\frac{\pi n^{2}}{m} \cos(2 \pi\delta
m) \nonumber\\
&&\qquad + \sin\frac{\pi n^{2}}{m} \sin(2 \pi\delta m)
\Big\},
\end{eqnarray}
where $g(x)$ is given by (\ref{rkky-semicont}).
Using the fact that
\begin{eqnarray}
\label{sum1}
\sum_{n=-m+1}^{m} \cos\frac{\pi n^{2}}{m}   &=&
\begin{cases}
0, & m=\text{odd} \\
\sqrt{2m}, & m=\text{even}
\end{cases}, \nonumber\\
&=& \sum_{n=-m+1}^{m}
\sin\frac{\pi n^{2}}{m},
\end{eqnarray}
we rewrite the RKKY correction as
\begin{equation}
\label{ex2}
 w_{\rm FM}\simeq -\frac{ R}{4}  \sum_{l=1}^{\infty} \frac{g(2l/R)}{l^{3/2}} \Big\{
 \cos(4 \pi\delta
l) +  \sin(4 \pi\delta l)
\Big\}.
\end{equation}

\begin{figure}[tb]
\includegraphics[width=0.15\textwidth]{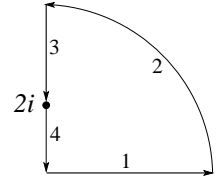}
\caption{
\label{fig:contour} The integration contour and subcontours used in
(\ref{subcont}). Bold lines denote cuts.
}
\end{figure}

Assuming that $R\gg1$ and $\delta\ll1$, and passing from the discrete sum to the continuum, we finally obtain
\begin{equation}
\label{w-fm-app}
w_{\rm FM}\simeq -\sqrt{\frac{R}{8}} A(2\pi\delta R) + 1,
\end{equation}
where
\begin{equation}
\label{int-fm}
A(Q) = \Re I(Q)  +\Im I(Q),\quad  I(Q)=\int_{0}^{\infty} dx \,\frac{g(x)}{x^{3/2}}e^{iQx},
\end{equation}
and the second term in (\ref{w-fm-app}) is the correction coming from the
fact that actually the lower integration limit in $I(Q)$ should be not
zero but $2/R$.
The integral $I(Q)$ above can be easily analyzed: integrating
$p(z)=\frac{g(z)}{z^{3/2}}e^{iQz}$ along the contour
shown in Fig.\ \ref{fig:contour} should obviously give zero. Assuming $Q>0$,  and sending the contour radius to infinity,
one can see that integrals over subcontours behave as
\begin{eqnarray}
\label{subcont}
&& \int_{C_{1}}p(z)dz\to I(Q),\quad \int_{C_{2}}p(z)dz \to 0, \nonumber\\
&&  \int_{C_{3}}p(z)dz\to -\pi\sqrt{2}(1-i)F(Q)+\frac{1+i}{\sqrt{2}}
P(Q),\nonumber\\
&& \int_{C_{4}}p(z)dz\to -(1+i) S(Q),
\end{eqnarray}
where we have used the following notation:
\begin{eqnarray}
\label{subint}
 F(Q)&=&\int_{2}^{\infty} dy\frac{e^{-Qy}}{y^{3/2}\sqrt{y^{2}-4}}, \nonumber\\
S(Q)&=&\int_{0}^{\pi/2} d\varphi \frac{\varphi}{\sin^{3/2}\varphi}
e^{-2Q\sin\varphi}, \\
 P(Q)&=&\int_{2}^{\infty} dy\frac{e^{-Qy}}{y^{3/2}\sqrt{y^{2}-4}} \ln \Big[
1-2y^{2}+\frac{y^{4}}{2} \nonumber \\
&+&y(\frac{y^{2}}{2}-1)\sqrt{y^{2}-4}
\Big].\nonumber
\end{eqnarray}
Thus, for $Q>0$, real and imaginary parts of $I(Q)$ are expressed via the above three
integrals as follows:
\begin{eqnarray}
\label{ReIm}
\Re I(Q) &=& \pi\sqrt{2} F(Q) -\frac{1}{\sqrt{2}} P(Q) +S(Q),\nonumber\\
 \Im I(Q) &=& -\pi\sqrt{2} F(Q) -\frac{1}{\sqrt{2}} P(Q) +S(Q).
\end{eqnarray}
To obtain $I(Q)$ for negative $Q$, one can notice that $\Re I(Q)$ is even in
$Q$, while $\Im I(Q)$ is odd.
The
combination $A(Q)= \Re I(Q)+ \Im I(Q)$ entering the energy correction (\ref{w-fm-app}) is thus expressed
as follows:
\begin{eqnarray}
\label{A-Q}
A(Q)= \begin{cases}
2S(Q) -\sqrt{2}P(Q), & Q>0\\
2\pi\sqrt{2}F(|Q|), & Q<0
\end{cases}.
\end{eqnarray}

Asymptotic behavior of the integrals $F$, $S$, and $P$ at small and large positive $Q$ can be
obtained in a standard way, here we just list the results:
\begin{equation}
\label{F-asymp}
F(Q)\simeq \begin{cases}
0.42-1.85Q, & Q\ll 1\\
\sqrt{\frac{\pi}{32Q}} e^{-2Q}, & Q\gg 1
\end{cases},
\end{equation}

\begin{equation}
\label{S-asymp}
S(Q)\simeq \begin{cases}
2.90-2.89Q, & Q\ll 1\\
\sqrt{\frac{\pi}{2Q}} \Big(1+  \frac{1}{8Q^{2}} \Big), & Q\gg 1
\end{cases},
\end{equation}

\begin{equation}
\label{P-asymp}
P(Q)\simeq \begin{cases}
1.44-15.73Q, & Q\ll 1\\
\frac{1}{\sqrt{2}Q} e^{-2Q}, & Q\gg 1
\end{cases}.
\end{equation}
After substituting all those expressions into (\ref{A-Q}),  one obtains the
asymptotics given by (\ref{A-asymp}) and resulting in the expressions (\ref{w-fm-asymp}).

\paragraph{Antiferromagnetic case.--}

 Setting $k_{F}$ in the vicinity of the first ``special point'', $k_{F}=\frac{\pi}{\sqrt{2}} (1-\delta)$, $|\delta|\ll 1$, one can see
that the main
contribution to the lattice sum in (\ref{rkky-semicont}) comes from the four equivalent
``cones'' around the directions $\vec{\Delta}=(\pm 1,\pm 1)$,
$\vec{\Delta}=(\pm1,\mp1)$.  The  contribution of one
``cone'' can be  approximately calculated as follows:  put
$\vec{\Delta}=(s+m,s-m)$, where $s$ and $m$ can be either both integer, or both half-integer,
with $|m| \ll s $, then $\eta_{\vec{\Delta}}=(-1)^{2s}$, inside the cosine $\cos(2k_{F}\Delta)$ one can expand $\Delta$ in $m$,
$\Delta\simeq s\sqrt{2}\big[1+m^{2}/(2s)\big]$,
and in the rest of the expression one can set $\Delta\simeq s\sqrt{2}$. This yields
\begin{eqnarray}
\label{ex1-af}
&& w_{\rm AFM} \simeq -\frac{ R}{2^{5/2}}
\sum_{s=\frac{1}{2},1,\frac{3}{2},\ldots}
\frac{g\left(\frac{s\sqrt{2}}{R}\right)}{s^{2}} \\
&&\quad\sum_{m=-s+1}^{s}
\Big\{
\cos\frac{\pi m^{2}}{s} \cos(2 \pi\delta
s)
 + \sin\frac{\pi m^{2}}{s} \sin(2 \pi\delta s)
\Big\}.\nonumber
\end{eqnarray}
Sums over $m$ can be readily performed: for integer $s$ they are given by Eq.\ (\ref{sum1}), and for half-integer
$s=l+\frac{1}{2}$ one has
\begin{eqnarray}
\label{sum2}
&&\sum_{m=-l+\frac{1}{2}}^{l+\frac{1}{2}} \cos\frac{\pi m^{2}}{l+\frac{1}{2}} =
\begin{cases}
\sqrt{2l+1}, & l=\text{odd} \\
0, & l=\text{even}
\end{cases},\\
&&\sum_{m=-l+\frac{1}{2}}^{l+\frac{1}{2}} \sin\frac{\pi m^{2}}{l+\frac{1}{2}} =
\begin{cases}
\sqrt{2l+1}, & l=\text{even} \\
0, & l=\text{odd}
\end{cases}.
\end{eqnarray}
Passing in the remaining sum over $s$ to the continuum, one can approximately express the RKKY
energy correction in terms of the same function  (\ref{int-fm}) introduced above in the
ferromagnetic case:
\begin{equation}
\label{w-afm-Q}
w_{\rm AFM}\simeq -\frac{\sqrt{R}}{2^{7/4}}A(Q=\pi\sqrt{2}\delta R) +1,
\end{equation}
where the origin of the second term is the same as in
Eq.\ (\ref{w-fm-app}), namely, the correction connected with the
nonzero lower integration limit when passing to the continuum.
Thus, the  behavior of the energy correction on the skyrmion radius $R$ for an
antiferromagnet  is qualitatively similar to that for a ferromagnet. The only
important difference is the change of the condition for the Fermi wave vector.

\section{``Hybrid'' formula for the lattice sum}
\label{app:hybrid}

Numerical computation of the fourfold lattice sum (\ref{rkky-2}) over
$\vec{r}$, $\vec{\Delta}$can be
a resource-intensive task for large lattices. ``Semi-continuum''
formula (\ref{rkky-semicont}), which converts the sum over $\vec{r}$
into an integral, simplifies this task considerably,
but introduces an uncontrollable approximation. To estimate the errors
of this approximation, one can undertake a ``hybrid'' approach,  using
the sum (\ref{rkky-2}) for $|\vec{r}|$ smaller than some radius
$r_{0}$, and passing to the continuum in $\vec{r}$
for $|\vec{r}|>r_{0}$. The corresponding expression is easily derived
and has the following form:
\begin{widetext}
\begin{eqnarray}
\label{hybrid}
\frac{\Delta E_{\rm RKKY}}{J'S^{2}} &=& -R^{2}\sum_{\vec{\Delta}} \eta_{\vec{\Delta}}\Delta^{2}
f(\vec{\Delta})
\left\{
 \sum_{|\vec{r}|<r_{0}} \frac{   (r^{2}+ R^{2}+
  \Delta^{2})}{(r^{2}+R^{2})
  [(r^{2}+R^{2}+\Delta^{2})^{2}-4(\vec{r}\cdot\vec{\Delta})^{2} ]}
 +\frac{\pi}{R^{2}}h\left(\frac{\Delta}{R},\frac{r_{0}}{R}\right) \right\}\nonumber,\\
 h(x,y)& =&\frac{1}{x\sqrt{4+x^{2}}}\ln\Big[
\frac{x(3+x^{2}-y^{2})+ \sqrt{4+x^{2}}\sqrt{(x^{2}+1)^{2}+y^{2}(2-2x^{2}+y^{2}) }}{(1+y^{2})(\sqrt{4+x^{2}}-x)}
\Big]
\end{eqnarray}
\end{widetext}


\end{document}